\documentclass[conference]{IEEEtran}
\IEEEoverridecommandlockouts

\usepackage[utf8]{inputenc} 
\usepackage[T1]{fontenc}
\usepackage{ifthen}
\usepackage{array}
\usepackage{mathrsfs}
\usepackage{amsthm}
\usepackage{multirow}
\usepackage{multicol}
\usepackage[dvipsnames]{xcolor}
\usepackage{pgfplots}
\pgfplotsset{compat=1.18}
\usepackage{comment}
\usepackage{braket}
\usepackage{stmaryrd}
\usepackage{mathtools}
\usepackage{bm}
\usepackage{amsmath,amssymb,amsfonts}
\usepackage{graphicx}
\usepackage{textcomp}
\usepackage[textsize=tiny]{todonotes}
\usepackage{tikz}
\usetikzlibrary{patterns}
\usepackage{booktabs}
\usepackage[caption=false,font=footnotesize]{subfig}

\newcommand{\sym}[1]{#1}         
\newcommand{\6}{\mathbf } 
\newcommand{\code}[1]{\mathscr{#1}} %

\makeatletter
\renewcommand*\env@matrix[1][*\c@MaxMatrixCols c]{%
  \hskip -\arraycolsep
  \let\@ifnextchar\new@ifnextchar
  \array{#1}}
\makeatother

\usepackage[ruled, vlined, linesnumbered]{algorithm2e}

\SetCommentSty{mycommfont}

\usepackage[noadjust]{cite} 

\theoremstyle{plain}

\newtheorem{Prop}{Proposition}
\newtheorem{Lem}{Lemma}
\newtheorem{Cor}{Corollary}

\newtheorem{Def}{Definition}

\theoremstyle{definition}
\newtheorem{Exa}{Example}

\theoremstyle{remark}

\newcommand{\rk}{\mathrm{rk}}

\newcommand{\rs}{\mathrm{row}}

\newcommand{\wt}[1]{\mathrm{wt}({#1})}

\usepackage[acronym,shortcuts]{glossaries}
\newacronym{CSS}{CSS}{Calderbank--Shor--Steane}
\newacronym{DPM}{DPM}{dyadic permutation matrix}
\newacronym{LER}{LER}{logical error rate}
\newacronym{MILP}{MILP}{mixed-integer linear program}
\newacronym{LDPC}{LDPC}{low-density parity-check}
\newacronym{QLDPC}{qLDPC}{quantum low-density parity-check}
\newacronym{BCH}{BCH}{Bose-Chaudhuri-Hocquenghem}
\newacronym{QC}{QC}{quasi-cyclic}
\newacronym{QC-LDPC}{QC-LDPC}{quasi-cyclic low-density parity-check}
\newacronym{QD}{QD}{quasi-dyadic}
\newacronym{QM}{QM}{quantum Margulis}
\newacronym{QD-LDPC}{QD-LDPC}{quasi-dyadic low-density parity-check}
\newacronym{BP}{BP}{belief propagation}
\newacronym{BP4}{BP$4$}{quaternary belief propagation}
\newacronym{CN}{CN}{check-node}
\newacronym{MS}{MS}{Min-Sum}
\newacronym{BP+OSD}{BP2+OSD}{belief propagation plus ordered-statistics decoding}
\newacronym{QECC}{QECC}{quantum error-correcting code}
\newacronym{QSC}{QSC}{quantum stabilizer code}
\newacronym{QEC}{QEC}{quantum error correction}
\newacronym{RM}{RM}{Reed-Muller}
\newacronym{MSD}{MSD}{magic state distillation}
\newacronym{GB}{GB}{generalized bicycle}
\newacronym{BB}{BB}{bivariate bicycle}
\newacronym{LP}{LP}{linear programming}
\newacronym{SD}{SD}{self-dual}
\newacronym{GHP}{GHP}{generalized hypergraph product}
\newacronym{DC}{DC}{dual-containing}
\newacronym{IS}{IS}{intersecting subsets}
\newacronym{HP}{HP}{hypergraph product}
\newacronym{LHCB}{LHCB}{left-hand conveyor belt}
\newacronym{GM}{GM}{generator matrix}
\newacronym{SD-XYZ}{SD-XYZ}{self-dual XYZ stabilizer}
\newacronym{LC}{LC}{local Clifford}
\newacronym{GRM}{GRM}{generalized Reed-Muller}
\newacronym
  [longplural={parity-check matrices},
   shortplural={PCMs}]
  {PCM}{PCM}{parity-check matrix}
\newacronym{AED}{AED}{automorphism ensemble decoder}

\pdfmapfile{+custom.map}

\usepackage{mathabx}
\usepackage{pifont}

\usepackage{booktabs}

\usepackage{array}

\usepackage[normalem]{ulem}
  \usepackage{url}

\theoremstyle{definition}

\theoremstyle{definition}
\newtheorem{remark}{Remark}

\begin{document}

\title{
Quantum Cyclic XYZ Stabilizer Codes
\thanks{The work of Alessio Baldelli was partially supported by Agenzia per la Cybersicurezza Nazionale (ACN) under the programme for promotion of XL cycle PhD research in cybersecurity (CUP I32B24001750005). 
The work of Francisco L\'azaro and Davide Orsucci is part of the HESC lighthouse project within Munich Quantum Valley initiative and is supported by the Bavarian state government with funds from the Hightech Agenda Bavaria.
}
}

\author{
\IEEEauthorblockN{
    Alessio Baldelli\IEEEauthorrefmark{2},
    Francisco Lázaro\IEEEauthorrefmark{3},
    Davide Orsucci\IEEEauthorrefmark{3},
    Massimo Battaglioni\IEEEauthorrefmark{2}
}
\IEEEauthorblockA{
\IEEEauthorrefmark{2}\emph{
    Department of Information Engineering,} Universit{\`a} Politecnica delle Marche, 
    60131 Ancona, Italy \\
    Email: \texttt{a.baldelli@pm.univpm.it}, \texttt{m.battaglioni@univpm.it} \\
}
\IEEEauthorblockA{
\IEEEauthorrefmark{3}\emph{
    Institute of Communications and Navigation,} German Aerospace Center (DLR), 
    82234 We{\ss}ling, Germany \\
    Email: \texttt{\{francisco.lazaroblasco, davide.orsucci\}@dlr.de}
}
\vspace{-10mm}
}

\maketitle

\begin{abstract}
    XYZ codes generalize the Calderbank--Shor--Steane (CSS) framework by adding $\sym{Y}$-type stabilizers to the usual $\sym{X}$- and $\sym{Z}$-type ones.
    We study XYZ codes whose three classical component codes are binary cyclic codes of odd length~$n$.
    Assigning disjoint sets of irreducible factors of $x^n-1$ to the three Pauli types fixes the number of encoded logical qubits and ensures that the row spaces of the three parity-check matrices intersect pairwise trivially.
    Then, a coprimality condition between the factors assigned to each type and the reciprocals of those assigned to the others guarantees the required orthogonality.
    Cyclotomic cosets give a complete and explicit description of the resulting family. 
    We show that, for odd~$n$, a cyclic stabilizer code admits an XYZ representation with cyclic component codes iff its stabilizer generators are invariant, up to a phase, under the qubit permutation $i\mapsto 2i\bmod n$.
    By fixing the length and the dimension of the code, the proposed cyclic XYZ codes attain the best known minimum distance of stabilizer codes, exceed the minimum distance of the underlying CSS codes, and even exceed the linear-programming upper bounds for general CSS codes.
\end{abstract}

\section{Introduction}
\label{sec:intro}

Within the family of stabilizer codes~\cite{gottesman1997stabilizercodesquantumerror},\ac{CSS} codes~\cite{Calderbank1996, Steane1996} are built using two classical codes satisfying an orthogonality condition, and their stabilizer generators consist solely of $\sym{X}$-type or $\sym{Z}$-type Pauli operators.
This structure brings several practical benefits, such as decoding $\sym{X}$- and $\sym{Z}$-type errors separately and implementing the logical CNOT gate transversally between two code blocks~\cite{gottesman1997stabilizercodesquantumerror}. 
However, the \ac{CSS} structure can be  more restrictive in terms of achievable \emph{minimum distance} $d$. For a given pair $(k, d)$, where $k$ is the number of encoded logical qubits, a \ac{CSS} code may require more physical qubits than a non-\ac{CSS} stabilizer code. 
For example, consider $(k, d) = (1, 3)$. 
The non-\ac{CSS} five-qubit code $\code{C} \llbracket 5, 1, 3 \rrbracket$~\cite{Laflamme_1996} uses only five physical qubits, whereas the smallest \ac{CSS} code with these parameters is the Steane code $\code{C} \llbracket 7, 1, 3 \rrbracket$~\cite{Steane1996}. Likewise, for fixed $(n,k)$, non-\ac{CSS} stabilizer codes can achieve larger minimum  distances than \ac{CSS} ones.

A natural way to go beyond \ac{CSS} codes, while retaining a construction based on classical codes, is to introduce $\sym{Y}$-type stabilizer generators.
This is the idea behind \emph{quantum XYZ stabilizer codes} (simply \emph{XYZ codes}), in which  the third Pauli type can increase the minimum distance by suppressing low-weight logical operators~\cite{xyz_codes}.

Cyclic codes are among the most studied families in classical coding theory: a single generator polynomial, or its set of roots, controls the code dimension and, through bounds such as the BCH bound, the minimum distance~\cite{macwilliams1977}.
In the quantum domain, cyclic and \ac{QC} codes have been widely exploited within the \ac{CSS} framework~\cite{sparse_McKay, Hagiwara2007, Kovalev_Pryadko, IBM_BBcodes}. More generally, cyclic stabilizer codes correspond to additive cyclic codes over $\mathbb F_4$~\cite{calderbank1998quantum, Huffman2007}, and have been studied extensively in~\cite{grassl1999bch, grassl2000cyclic, Bierbrauer2000twisted, aly2007bch, ketkar2006}. 
As shown in~\cite{Grassl_table, GrasslCyc}, many of the best known stabilizer codes arise from cyclic constructions.

In this work, we specialize the XYZ framework of~\cite{xyz_codes} to cyclic component codes. 
For odd $n$, we assign each irreducible factor of $x^n-1$ to at most one Pauli type and call the resulting codes \emph{cyclic XYZ codes}.
This fixes the number of logical qubits and makes the row spaces of the three \acp{PCM} intersect pairwise trivially, while orthogonality requires that each cyclotomic coset and its negative contribute to at most one Pauli type. 
This yields a complete and explicit description of the family. Relative to the theory of additive cyclic codes over~$\mathbb F_4$~\cite{calderbank1998quantum, Huffman2007}, our structural results are twofold. 
First, among cyclic stabilizer codes of odd length, an XYZ representation with cyclic component codes exists if and only if the code is invariant under the qubit permutation $i\mapsto 2i\bmod n$. Our construction then yields precisely those codes for which all three Pauli types occur and the row spaces of the three \acp{PCM} intersect pairwise trivially.
Second, we uncover a Pauli-symmetric subclass in which a suitable permutation of the qubits, combined with the cyclic exchange $X\to Y\to Z\to X$, leaves the code unchanged.

Notably, the proposed codes exceed the minimum distance of their underlying \ac{CSS} codes and attain the best known minimum distance of stabilizer codes with the same length and dimension~\cite{Grassl_table}. 
Moreover, their minimum distance exceed the \ac{LP} upper bounds for \ac{CSS} codes with the same length and dimension~\cite{Delsarte1973, Shor1997_MacWilliams, Rains1998_enumerators, Ashikhmin1999_bounds}.

\section{Notation and Background}
\label{sec:preli}

Row vectors are denoted by bold lowercase letters, while matrices are represented by bold uppercase letters, e.g., $\6a$ and $\6A$. The symbol $\60$ denotes the all-zero vector or matrix of the appropriate size. Unless otherwise stated, vector and matrix operations are over the binary field $\mathbb{F}_2$. We denote by $\mathbb{Z}_n$ the ring of integers modulo $n$, with unit group $\mathbb{Z}_n^{*}$. We write $\rk(\6A)$ and $\rs(\6A)$ for the rank and row space of $\6A$, respectively, and $\wt{\6a}$ for the Hamming weight of $\6a$.

\subsection{Classical Linear and Cyclic Codes}
\label{subsec:cyclic_codes}

A binary linear code $\code{C}[n,k,\delta]\subseteq\mathbb{F}_2^n$ of length $n$, dimension $k$, and minimum distance $\delta$ is the null space $\{\6c \mid \6H\6c^\top=\60\}$ of a \ac{PCM} $\6H$ with $\rk(\6H)=n-k$. Its dual code is $\code{C}^\perp=\rs(\6H)$.

A classical linear code is \emph{cyclic} if it is invariant under the cyclic shift of coordinates. Identifying $(v_0,\dots,v_{n-1})\in\mathbb{F}_2^n$ with $v(x)=\sum_{i=0}^{n-1}v_ix^i$ in $\mathcal{R}_n=\mathbb{F}_2[x]/(x^n-1)$, the cyclic shift becomes multiplication by $x$. 
The cyclic codes of length $n$ are the ideals $\langle a\rangle$ with $a\mid x^n-1$, with $k = n-\deg a$~\cite{macwilliams1977}.
The \emph{reciprocal} of $a$ is $\tilde a$, defined by  $\tilde a(x)=x^{\deg a}a(1/x)$, i.e., $a$ with its coefficients in reverse order; $a$ is \emph{self-reciprocal} if $\tilde a=a$.
Then $\langle a\rangle\perp\langle b\rangle$ iff $(x^n-1)\mid a\tilde b$, and $\langle a\rangle$ is \emph{reversible}, i.e., invariant under reversing the order of the coordinates, iff $a$ is self-reciprocal~\cite{macwilliams1977}. Furthermore, for every divisor $a$ of $x^n-1$, the dual of $\langle a\rangle$ is $\langle(x^n-1)/\tilde a\rangle$.

For odd $n$, let $\alpha$ be a primitive $n$-th root of unity in an extension field of $\mathbb{F}_2$, so that $x^n-1=\prod_{i\in\mathbb{Z}_n}(x-\alpha^i)$.
The roots of any factor of $x^n-1$ in $\mathbb{F}_2[x]$ are closed under doubling of the exponents, so the irreducible factors of $x^n-1$ are the minimal polynomials $M_s(x)=\prod_{i\in\mathcal{C}_s}(x-\alpha^i)$, indexed by the \emph{cyclotomic cosets} $\mathcal{C}_s=\{s,2s,4s,\dots\}\subseteq\mathbb{Z}_n$. A cyclic code $\langle a\rangle$ is thus specified by its \emph{defining set} $\mathcal{T}=\{i\in\mathbb{Z}_n \mid a(\alpha^i)=0\}$, a union of cyclotomic cosets with $|\mathcal{T}|=\deg a$.
The reciprocal $\tilde a$ has defining set $-\mathcal{T}$, so $a$ is self-reciprocal iff $\mathcal{T}=-\mathcal{T}$.  
Since $x^n-1$ is squarefree, $\mathcal{R}_n$ is the direct sum of the minimal ideals $\mathcal{I}_s=\langle(x^n-1)/M_s\rangle$, each a field whose identity $e_s$ is an idempotent, with $\sum_s e_s=1$ and $e_se_{s'}=0$ for $s\neq s'$, and $\langle(x^n-1)/f\rangle=\bigoplus_{s:\,M_s\mid f}\mathcal{I}_s$ for every $f\mid x^n-1$~\cite[Ch.~8, Section~3, Thms.~7 and~9]{macwilliams1977}.
For $t\in\mathbb{Z}_n^{*}$, let $\mu_t$ denote the \emph{multiplier} $v(x)\mapsto v(x^t)$, which permutes the coordinates as $i\mapsto ti\bmod n$ and maps the cyclic code with defining set $\mathcal{T}$ to the one with defining set $t^{-1}\mathcal{T}$; hence, cyclic codes related by a multiplier are permutation equivalent.

\subsection{Quantum Stabilizer Codes and XYZ Codes}
\label{subsec:stabilizer_codes}
We represent an $n$-qubit Pauli operator, up to phase, by $[\6a \, | \, \6b]\in\mathbb{F}_2^{2n}$, where $\6a$ and $\6b$ are its $\sym{X}$ and $\sym{Z}$ parts, so that $\sym{X}$-, $\sym{Z}$- and $\sym{Y}$-type operators read $[\6a\,|\,\60]$, $[\60\,|\,\6b]$ and $[\6c\,|\,\6c]$, respectively; a stabilizer code is then described by a \ac{PCM} with $2n$ columns whose rows represent Pauli operators and are pairwise orthogonal under the symplectic product $\langle[\6a\,|\,\6b],[\6a'\,|\,\6b']\rangle=\6a\cdot\6b'+\6b\cdot\6a'$. 
Let $\mathcal{S} \subseteq \mathbb{F}_2^{2n}$ be the row space of the \ac{PCM}, and call the code \emph{cyclic} (resp.\ \emph{reversible}) if $\mathcal{S}$ is invariant under the
simultaneous cyclic shift (resp.\ reversal of the order of the coordinates) of both halves.
$\mathcal{S}$ is also known as stabilizer group of the stabilizer code.
The minimum distance $d$ is the smallest weight, i.e., number of non-identity factors, of a nontrivial logical operator, represented by an element of $\mathcal{S}^{\perp_s}\setminus\mathcal{S}$, where $\perp_s$ denotes the dual with respect to the symplectic product.  A nonzero element of $\mathcal{S}$ is called \emph{pure $\sigma$-type} if its corresponding Pauli operator contains only $\sigma$ and identity factors, where $\sigma\in\{\sym{X},\sym{Y},\sym{Z}\}$.

\begin{Def}[XYZ codes~\cite{xyz_codes}]
\label{def:XYZ_codes}
    Let $\code{C}_\sigma[n,k_\sigma,\delta_\sigma]$, $\sigma\in\{\sym{X},\sym{Y},\sym{Z}\}$, be 
    codes with \acp{PCM} $\6H_\sigma\in\mathbb{F}_2^{m_\sigma\times n}$ such that
    \begin{align} \label{eq:prod_XYZ}
        \6H_{\mathrm{X}} \6H_{\mathrm{Y}}^\top = \60, \quad \6H_{\mathrm{Y}} \6H_{\mathrm{Z}}^\top = \60, \quad \6H_{\mathrm{Z}} \6H_{\mathrm{X}}^\top = \60.
    \end{align}
    The \emph{XYZ code} $\code{C} \llbracket n,k,d\rrbracket$ is a stabilizer code with \ac{PCM}
    \begin{align} 
    \label{eq:PCM_XYZ}
        \6H_{\mathrm{XYZ}} :=
        \begin{bmatrix} [c|c]
            \6H_{\mathrm{X}} & \60 \\
            \60   & \6H_{\mathrm{Z}} \\
            \6H_{\mathrm{Y}} & \6H_{\mathrm{Y}}
        \end{bmatrix}
        \in \mathbb{F}_2^{(m_{\mathrm{X}} + m_{\mathrm{Y}} + m_{\mathrm{Z}}) \times 2n}.
    \end{align}
   The code encodes $k=n-\rk(\6H_{\mathrm{XYZ}})$ logical qubits.
\end{Def}

For each Pauli type $\sigma$, the \emph{check space} $\mathcal{W}_\sigma\subseteq\mathbb{F}_2^n$ is
\begin{align*}
\mathcal{W}_{\mathrm{X}} = \{\6v\mid[\6v\,|\,\60]\in\mathcal{S}\} &, \quad 
\mathcal{W}_{\mathrm{Z}} = \{\6v\mid[\60\,|\,\6v]\in\mathcal{S}\}, \\
\mathcal{W}_{\mathrm{Y}} = & 
\{\6v\mid[\6v\,|\,\6v]\in\mathcal{S}\}.
\end{align*}
For a \ac{PCM} of the form~\eqref{eq:PCM_XYZ}, every row of $\6H_\sigma$ gives a pure $\sigma$-type element, so $\rs(\6H_\sigma)\subseteq\mathcal{W}_\sigma$. The next lemma gives a simple condition for equality.

\begin{Lem}[Trivially intersecting blocks~\cite{xyz_codes}]
\label{lem:trivial_int}
    Consider an XYZ code whose row spaces $\rs(\6H_{\mathrm X})$, $\rs(\6H_{\mathrm Y})$ and $\rs(\6H_{\mathrm Z})$ intersect pairwise trivially. 
    Then
    \begin{enumerate}
        \item[$(i)$] $\rk(\6H_{\mathrm{XYZ}})=\rk(\6H_{\mathrm X})+\rk(\6H_{\mathrm Y})+\rk(\6H_{\mathrm Z})$;
        \item[$(ii)$] $\mathcal{W}_\sigma=\rs(\6H_\sigma)$ $\forall \, \sigma$; and $\mathcal{W}_\sigma\cap\mathcal{W}_\tau=\{\60\}$ for $\sigma\neq\tau$.
    \end{enumerate}
\end{Lem}

\begin{IEEEproof}
    Every element of $\mathcal{S}$ is $[\6a+\6c\,|\,\6b+\6c]$ with $\6a\in\rs(\6H_{\mathrm X})$, $\6b\in\rs(\6H_{\mathrm Z})$ and $\6c\in\rs(\6H_{\mathrm Y})$. If it is $\60$, then $\6a=\6c=\6b$, so $\6c\in\rs(\6H_{\mathrm X})\cap\rs(\6H_{\mathrm Y})=\{\60\}$, which gives $(i)$. It is pure $\sym{X}$-type iff $\6b=\6c\in\rs(\6H_{\mathrm Z})\cap\rs(\6H_{\mathrm Y})=\{\60\}$, i.e., iff it equals $[\6a\,|\,\60]$.
    The $\sym{Z}$ and $\sym{Y}$ types follow in the same way from $\rs(\6H_{\mathrm X})\cap\rs(\6H_{\mathrm Y})=\{\60\}$ and $\rs(\6H_{\mathrm X})\cap\rs(\6H_{\mathrm Z})=\{\60\}$, which gives $(ii)$.
\end{IEEEproof}

Deleting one of the three blocks of~\eqref{eq:PCM_XYZ} leaves a stabilizer code with two Pauli types, which is a \ac{CSS} code up to a transversal relabeling of the Pauli types; we call these the three \emph{underlying \ac{CSS} codes} of the XYZ code.
An XYZ code may still be a \ac{CSS} code in disguise: its stabilizer group may be generated by only two Pauli types, possibly after applying a single-qubit Clifford gate to each qubit~\cite{xyz_codes}.

\begin{Def}[Genuine codes~\cite{xyz_codes}]
\label{def:genuine_XYZ}
    An XYZ code is \emph{genuine} if it is not \ac{LC} equivalent to a \ac{CSS} code.
\end{Def}

\section{Cyclic XYZ Codes}
\label{sec:cyclic_xyz}
 
We construct cyclic XYZ codes from cyclic classical codes, choosing each \ac{PCM} $\6H_\sigma$ in~\eqref{eq:PCM_XYZ} to be circulant.

\begin{Def}[Cyclic XYZ codes]
\label{def:cyclic_XYZ}
    Let $n$ be odd and let
    \begin{equation}
    \label{eq:cyc_split}
        x^n-1=f_{\mathrm{X}}(x)f_{\mathrm{Y}}(x)f_{\mathrm{Z}}(x)g(x)
    \end{equation}
    be a factorization over $\mathbb{F}_2$ with $ f_\sigma\neq 1$  
    $\forall \, \sigma\in\{\sym{X},\sym{Y},\sym{Z}\}$ and
    \begin{equation}
    \label{eq:cond_orth}
        \gcd(f_\sigma,\tilde f_\tau)=1 \quad \text{for all } \sigma\neq\tau.
    \end{equation}
    Let $h_\sigma(x)=(x^n-1)/f_\sigma(x)$, and let $\6H_\sigma$ be the $n\times n$ circulant matrix whose rows are the cyclic shifts of the coefficient vector of $h_\sigma$. 
    The \emph{cyclic XYZ code} is denoted as $\code{C}_{\mathrm{cyc}}\llbracket n,k,d\rrbracket$.
\end{Def}

Since $n$ is odd, $x^n-1$ is squarefree, so the factors in~\eqref{eq:cyc_split} are pairwise coprime, that is
\begin{equation}
\label{eq:cond_coprime}
    \gcd(f_\sigma,f_\tau)=1 \quad \text{for all } \sigma\neq\tau.
\end{equation}
By Section~\ref{subsec:cyclic_codes}, $\rs(\6H_\sigma)=\langle h_\sigma\rangle$ has dimension $\deg f_\sigma$, and the component code is $\code{C}_\sigma=\langle h_\sigma\rangle^{\perp}=\langle\tilde f_\sigma\rangle$. 
The condition $f_\sigma\neq1$ ensures that all three Pauli types occur.
Conversely, if $f_\sigma = 1$, only two Pauli types remain and the code is not genuine (Definition~\ref{def:genuine_XYZ}).

\begin{Prop}[Orthogonality and dimension]
\label{prop:cyc_params}
   The matrices of Definition~\ref{def:cyclic_XYZ} satisfy the orthogonality conditions in~\eqref{eq:prod_XYZ} and define a cyclic XYZ code $\code{C}_{\mathrm{cyc}}\llbracket n,k,d\rrbracket$ with $k=\deg g$, whose check spaces $\mathcal{W}_\sigma=\langle h_\sigma\rangle$ intersect pairwise trivially.
\end{Prop}

\begin{IEEEproof}    
    Since $x^n-1=f_\tau h_\tau$ and $x^n-1$ is self-reciprocal, taking reciprocals gives $\tilde h_\tau=(x^n-1)/\tilde f_\tau$. For $\sigma\neq\tau$, $f_\sigma$ and $\tilde f_\tau$ are coprime divisors of $x^n-1$ by~\eqref{eq:cond_orth}, so 
    \begin{equation*}
          h_\sigma\tilde h_\tau=(x^n-1)\,\frac{x^n-1}{f_\sigma\tilde f_\tau}\equiv0\pmod{x^n-1}.
    \end{equation*}
    By the orthogonality criterion of Section~\ref{subsec:cyclic_codes}, $\langle h_\sigma\rangle\perp\langle h_\tau\rangle$, i.e., $\6H_\sigma\6H_\tau^\top=\60$.
    Moreover, by Section~\ref{subsec:cyclic_codes}, $\langle h_\sigma\rangle$ is the direct sum of the minimal ideals $\mathcal{I}_s$ with $M_s\mid f_\sigma$, and by~\eqref{eq:cond_coprime} no $M_s$ divides two different $f_\sigma$. Hence, the row spaces $\langle h_\sigma\rangle$ intersect pairwise trivially, and Lemma~\ref{lem:trivial_int} gives $\mathcal{W}_\sigma=\langle h_\sigma\rangle$ and, by~\eqref{eq:cyc_split}, $\rk(\6H_{\mathrm{XYZ}})=\sum_\sigma\deg f_\sigma=n-\deg g$, i.e., $k=\deg g$.
\end{IEEEproof}

\subsection{Construction from Cyclotomic Cosets}
\label{sec:cyclotomic}

By Definition~\ref{def:cyclic_XYZ}, a cyclic XYZ code is determined by $f_{\mathrm X}$, $f_{\mathrm Y}$ and $f_{\mathrm Z}$. Each $f_\sigma$ divides $x^n-1$, so it can be represented by its defining set $\mathcal{T}_\sigma$, a union of cyclotomic cosets with $|\mathcal{T}_\sigma|=\deg f_\sigma$. We now translate conditions~\eqref{eq:cond_orth} and~\eqref{eq:cond_coprime} into a rule on these sets. This rule lists all cyclic XYZ codes of length $n$ explicitly and shows which of them are reversible.

As discussed above, $f_\sigma$ is the product of the minimal polynomials $M_s$ with $\mathcal{C}_s\subseteq\mathcal{T}_\sigma$. Condition~\eqref{eq:cond_coprime} says that no $M_s$ is used by two Pauli types. The reciprocal $\tilde M_s$ has defining set $-\mathcal{C}_s=\mathcal{C}_{-s}$, i.e., $\tilde M_s=M_{-s}$; hence, condition~\eqref{eq:cond_orth} says that, if $\sigma$ uses $M_s$, no other type uses $M_{-s}$. In short, the two factors $M_s$ and $M_{-s}$ belong to at most one Pauli type, which may use either of them or both.

In particular, these two factors correspond to the \emph{negation orbit} $\mathcal{O}_s=\mathcal{C}_s\cup\mathcal{C}_{-s}$. 
It is one coset if $M_s$ is self-reciprocal, i.e., $\mathcal{C}_s=\mathcal{C}_{-s}$ (\emph{self-negating} orbit), and of two cosets otherwise (\emph{paired} orbit). 
For prime $n$, this distinction is determined by whether $-1$ is a power of $2$ modulo $n$.. 
Thus, \eqref{eq:cond_orth} and~\eqref{eq:cond_coprime} hold iff every orbit meets at most one of the defining sets:
\begin{equation}
\label{eq:orbit_rule}
    \big|\{\sigma \, | \, \mathcal{O}_s\cap\mathcal{T}_\sigma\neq\varnothing\}\big|\leq1 \quad \text{for every orbit } \mathcal{O}_s.
\end{equation}
Within an orbit that meets $\mathcal{T}_\sigma$,  $\mathcal{T}_\sigma$ contains the whole orbit if it is self-negating, and $\mathcal{C}_s$, $\mathcal{C}_{-s}$ or both if it is paired. This yields the following construction.

\begin{Prop}[Cyclotomic construction]
\label{prop:cyc_construction}
    Let $n$ be odd, and let $\mathcal{O}_{\mathrm{X}},\mathcal{O}_{\mathrm{Y}},\mathcal{O}_{\mathrm{Z}}$ be non-empty pairwise disjoint sets of negation orbits. For each $\sigma$ and each orbit $\mathcal{O}_s\in\Omega_\sigma$, choose
    \begin{equation}
    \label{eq:coset_choice}
      \mathcal{T}_{\sigma,s} =
      \begin{cases}
        \mathcal{O}_s, & \text{if } \mathcal{O}_s \text{ is self-negating}, \\[2pt]
        \mathcal{C}_s,\ \mathcal{C}_{-s},\ \text{or}\ \mathcal{O}_s, & \text{if } \mathcal{O}_s \text{ is paired},
      \end{cases}
    \end{equation}
    and let $\mathcal{T}_\sigma=\bigcup_{\mathcal{O}_s\in\Omega_\sigma}\mathcal{T}_{\sigma,s}$. Then the polynomials $f_\sigma(x)=\prod_{i\in\mathcal{T}_\sigma}(x-\alpha^i)$ satisfy the conditions of Definition~\ref{def:cyclic_XYZ}, and the resulting cyclic XYZ code has dimension
    \begin{equation}
    \label{eq:k_cosets}
        k = n - |\mathcal{T}_{\mathrm{X}}| - |\mathcal{T}_{\mathrm{Y}}| - |\mathcal{T}_{\mathrm{Z}}|.
    \end{equation}
\end{Prop}

\begin{IEEEproof}
   Each $\mathcal{T}_\sigma$ is a union of cosets, so $f_\sigma$ is a product of minimal polynomials, hence a divisor of $x^n-1$ in $\mathbb{F}_2[x]$, with $\deg f_\sigma=|\mathcal{T}_\sigma|\geq1$, as $\Omega_\sigma\neq\varnothing$.  
   Moreover, $\mathcal{T}_\sigma$ meets exactly the orbits in $\Omega_\sigma$, so~\eqref{eq:orbit_rule} holds because the sets $\Omega_\sigma$ are pairwise disjoint. 
   Hence, \eqref{eq:cond_orth} and~\eqref{eq:cond_coprime} hold; in particular,  $f_\sigma$ are pairwise coprime, so their product divides $x^n-1$ and defines $g$. By Proposition~\ref{prop:cyc_params}, $k=\deg g=n-\sum_\sigma\deg f_\sigma$, which is~\eqref{eq:k_cosets}.
\end{IEEEproof}

\begin{Cor}[Converse construction]
\label{cor:cyc_converse}
    Every cyclic XYZ code of Definition~\ref{def:cyclic_XYZ} arises from Proposition~\ref{prop:cyc_construction}, with $\mathcal{T}_\sigma$ the defining set of $f_\sigma$ and $\Omega_\sigma$ the set of orbits that meet $\mathcal{T}_\sigma$.
\end{Cor}
\begin{IEEEproof}
   The defining sets $\mathcal{T}_\sigma$ satisfy~\eqref{eq:orbit_rule}, so no orbit meets two different $\mathcal{T}_\sigma$, and the sets $\Omega_\sigma$ are pairwise disjoint. Within each orbit $\mathcal{O}_s\in\Omega_\sigma$, the set $\mathcal{T}_\sigma\cap\mathcal{O}_s$ is a nonempty union of the cosets of $\mathcal{O}_s$, i.e., one of the choices in~\eqref{eq:coset_choice}. Their union is $\mathcal{T}_\sigma$, so these choices give back $f_\sigma$, and $\Omega_\sigma\neq\varnothing$ as $f_\sigma\neq1$.
\end{IEEEproof}

Whether a paired orbit is split, i.e., only one of its two cosets is kept, decides reversibility. Reversible codes are those with self-reciprocal $f_\sigma$, for which~\eqref{eq:cond_orth} already follows from~\eqref{eq:cond_coprime}; splitting gives up this symmetry but allows more values of $k$.

\begin{Cor}[Reversible construction]
\label{cor:rev_construction}
    The cyclic XYZ code of Proposition~\ref{prop:cyc_construction} is reversible iff no paired orbit is split, i.e., iff $\mathcal{T}_{\sigma,s}=\mathcal{O}_s$ for every $\mathcal{O}_s\in\Omega_\sigma$ and every $\sigma \in \{ \sym{X}, \sym{Y}, \sym{Z} \}$.
\end{Cor}

\begin{IEEEproof}
   Coordinate reversal $i\mapsto n-1-i$ is the multiplier $\mu_{-1}$ followed by a cyclic shift, so the cyclic $\mathcal{S}$ is reversible iff it is invariant under $\mu_{-1}$ applied to both halves. 
   This map preserves Pauli types and maps the cyclic code with defining set $\mathcal{D}$ to the one with defining set $-\mathcal{D}$. 
   As $\mathcal{S}$ is spanned by its pure elements, it is reversible iff every $\mathcal{W}_\sigma$ is. 
   By Proposition~\ref{prop:cyc_params}, $\mathcal{W}_\sigma=\langle h_\sigma\rangle$ has defining set $\mathbb{Z}_n\setminus\mathcal{T}_\sigma$; hence,  $\mathcal{S}$ is reversible iff $\mathcal{T}_\sigma=-\mathcal{T}_\sigma$ for every $\sigma$. 
   As the orbits are disjoint and closed under negation, this holds iff $\mathcal{T}_{\sigma,s}=-\mathcal{T}_{\sigma,s}$ for every $\mathcal{O}_s\in\Omega_\sigma$. 
   By~\eqref{eq:coset_choice}, $\mathcal{T}_{\sigma,s}$ is either the whole orbit $\mathcal{O}_s$ or, for a paired orbit, one of its two cosets $\mathcal{C}_s$ and $\mathcal{C}_{-s}$. Only the former is closed under negation, since $-\mathcal{C}_s=\mathcal{C}_{-s}\neq\mathcal{C}_s$ for a paired orbit.
\end{IEEEproof}

\subsection{Pauli-Symmetric Codes}
\label{sec:pauli_cyclic}

Proposition~\ref{prop:cyc_construction} treats the three Pauli types independently.
We now impose that they be related by a single multiplier, which forces the component codes to be permutation equivalent and endows the code with a symmetry permuting the Pauli types.

\begin{Def}[Pauli-symmetric XYZ codes]
\label{def:pauli_cyclic}
Let $t\in\mathbb{Z}_n^{*}$ satisfy  $t^3\equiv2^j\pmod n$ for some $j$, and let $\mathcal{T}_{\mathrm{X}}$ be a nonempty union of cyclotomic cosets modulo $n$. Let $\mathcal{T}_{\mathrm{Z}}=t\mathcal{T}_{\mathrm{X}}$ and $\mathcal{T}_{\mathrm{Y}}=t\mathcal{T}_{\mathrm{Z}}$; since $t^3\mathcal{T}_{\mathrm{X}}=2^j\mathcal{T}_{\mathrm{X}}=\mathcal{T}_{\mathrm{X}}$, we have
\begin{equation}
\label{eq:pauli_cyclic}
    \mathcal{T}_{\mathrm{X}}=t\mathcal{T}_{\mathrm{Y}},
    \qquad
    \mathcal{T}_{\mathrm{Y}}=t\mathcal{T}_{\mathrm{Z}},
    \qquad
    \mathcal{T}_{\mathrm{Z}}=t\mathcal{T}_{\mathrm{X}},
\end{equation}
and assume that these sets satisfy~\eqref{eq:orbit_rule}. The resulting cyclic XYZ code is called \emph{Pauli-symmetric}.
\end{Def}

The following proposition shows that this symmetry of the defining sets induces a symmetry of $\mathcal{S}$.

\begin{Prop}[Pauli symmetry]
\label{prop:pauli_symmetry}
For a Pauli-symmetric XYZ code, the multiplier $\mu_t$ followed by the transversal Clifford mapping $\sym{X}\to\sym{Y}\to\sym{Z}\to\sym{X}$ leaves $\mathcal{S}$ invariant. Moreover, $M_0\mid g$; in particular, $k\geq1$.
\end{Prop}
\begin{IEEEproof}
    By Proposition~\ref{prop:cyc_params}, $\mathcal{W}_\sigma=\langle h_\sigma\rangle$, whose defining set is $\mathbb{Z}_n\setminus\mathcal{T}_\sigma$. The action of multipliers described in Section~\ref{subsec:cyclic_codes}, together with~\eqref{eq:pauli_cyclic}, therefore gives
    \[
    \begin{aligned}
        \mu_t(\mathcal{W}_{\mathrm{X}}) = \mathcal{W}_{\mathrm{Y}}, \quad 
        \mu_t(\mathcal{W}_{\mathrm{Y}}) = \mathcal{W}_{\mathrm{Z}}, \quad     
        \mu_t(\mathcal{W}_{\mathrm{Z}}) = \mathcal{W}_{\mathrm{X}}.
    \end{aligned}
    \]
    Following $\mu_t$ by the transversal Clifford cycle thus maps pure $\sym{X}$-, $\sym{Y}$- and $\sym{Z}$-type elements of $\mathcal{S}$ onto pure $\sym{Y}$-, $\sym{Z}$- and $\sym{X}$-type elements, respectively. Since these elements span $\mathcal{S}$, the combined operation leaves $\mathcal{S}$ invariant.
    
    The Clifford cycle can be implemented without sign changes because it preserves the Pauli multiplication relations. 
    With this choice, and with all pure-type generators taken with phase $+1$, a consistent choice by Lemma~\ref{lem:trivial_int}$(i)$, the combined operation maps the stabilizer group onto itself, including its phases, and hence preserves the code space.
    
    Finally, since $t\cdot0=0$,~\eqref{eq:pauli_cyclic} and~\eqref{eq:orbit_rule} imply $0\notin\mathcal{T}_\sigma$ for every $\sigma$. 
    So, $M_0 \mid g$ and $k = \deg g \geq 1$.
\end{IEEEproof}

\subsection{Characterization by Symmetry}\label{subsec: frobenius}

Binary cyclic codes are invariant not only under the cyclic shift but also under the \emph{Frobenius multiplier} $\mu_2$, since $v(x^2) = v(x)^2$ over $\mathbb{F}_2$~\cite{macwilliams1977}. 
On a stabilizer code, $\mu_2$ acts on $\mathcal{S}$ by permuting both halves, and cyclic stabilizer codes are in general not invariant under $\mu_2$. Under the map $[\6a\,|\,\6b]\mapsto\omega\6a+\bar\omega\6b$ of~\cite{calderbank1998quantum}, $\mathcal{S}$ is an additive cyclic code over $\mathbb{F}_4$, whose canonical decomposition for odd length is given in~\cite{Huffman2007}. 
We rederive it in the binary representation and use it to show that, among the additive cyclic codes of~\cite{calderbank1998quantum}, invariance under $\mu_2$ identifies exactly those with an XYZ representation with cyclic component codes.Corollary~\ref{cor:frobenius_completeness} then identifies our cyclic XYZ codes among them.

\begin{Prop}[Frobenius characterization]
\label{prop:frobenius}
Let $\mathcal{S}$ be the stabilizer group of a cyclic stabilizer code of odd length $n$.  
The following are equivalent:
    \begin{enumerate}
        \item[$(i)$] $\mathcal{S}$ is invariant under $\mu_2$;
        \item[$(ii)$] $\mathcal{S}$ is generated by pure $\sym{X}$-, $\sym{Y}$- and $\sym{Z}$-type elements;
        \item[$(iii)$] $\mathcal{S}$ is the stabilizer group of an XYZ code whose component codes are cyclic.
    \end{enumerate}
\end{Prop}

\begin{IEEEproof}
    Let $\mathcal{I}_s$ and $e_s$ be as in Section~\ref{subsec:cyclic_codes}. We identify $[\6a\,|\,\6b]$ with $[a\,|\,b]\in\mathcal{R}_n\times\mathcal{R}_n$, write $p [a \, | \, b] = [pa\,|\,pb]$ for $p\in\mathcal{R}_n$, and call any $[a\,|\,b]\neq\60$ pure if $b=\60$, $a=\60$ or $a=b$. Since $\mathcal{S}$ is linear and invariant under multiplication by $x$, $p\mathcal{S}\subseteq\mathcal{S}$ for every $p\in\mathcal{R}_n$. Hence, $\mathcal{S}_s=e_s\mathcal{S}\subseteq\mathcal{S}\cap(\mathcal{I}_s\times\mathcal{I}_s)$, with equality as $e_s$ is the identity of $\mathcal{I}_s$, so $\mathcal{S}_s$ is a vector space over the field $\mathcal{I}_s$, and $\sum_se_s=1$ gives $\mathcal{S}=\bigoplus_s\mathcal{S}_s$. As $\mathcal{I}_s\times\mathcal{I}_s$ has dimension two over $\mathcal{I}_s$, each $\mathcal{S}_s$ is $\{\60\}$, $\mathcal{I}_s\times\mathcal{I}_s$, or a line $\mathcal{I}_sw=\{[cu\,|\,cv]\mid c\in\mathcal{I}_s\}$ with $w=[u\,|\,v]\neq\60$.

    Let $\mathcal{L}_{s,\sigma}$ be the line formed by $\60$ and the pure $\sigma$-type elements of $\mathcal{I}_s\times\mathcal{I}_s$, e.g., $\mathcal{L}_{s,\mathrm{X}}=\mathcal{I}_s[e_s\,|\,\60]$. Since multiplication by $p\in\mathcal{R}_n$ preserves Pauli types, a line $\mathcal{I}_sw$ contains a pure element iff $w$ is pure, i.e., iff $\mathcal{I}_sw=\mathcal{L}_{s,\sigma}$ for some $\sigma$. Indeed, if $cw$ is pure, then $c\neq\60$ and $w=c^{-1}(cw)$ is pure, while if $w$ is pure $\sigma$-type, its nonzero entries are invertible in $\mathcal{I}_s$, so $\mathcal{I}_sw=\mathcal{L}_{s,\sigma}$. We show that $(i)$ and $(ii)$ are both equivalent to requiring, for every $s$,
    \begin{equation}
    \label{eq:frob_lines}
        \dim_{\mathcal{I}_s}\mathcal{S}_s=1 \ \Rightarrow\ \mathcal{S}_s=\mathcal{L}_{s,\sigma} \text{ for some } \sigma.
    \end{equation}
    If~\eqref{eq:frob_lines} holds, every $\mathcal{S}_s$ is spanned by pure elements, as $\mathcal{I}_s\times\mathcal{I}_s=\mathcal{L}_{s,\mathrm{X}}\oplus\mathcal{L}_{s,\mathrm{Z}}$, and so is $\mathcal{S}=\bigoplus_s\mathcal{S}_s$. Conversely, if $\mathcal{S}$ is spanned by pure elements $g_j$, then $\mathcal{S}_s=e_s\mathcal{S}$ is spanned by the $e_sg_j$, which are pure or zero; hence, a line $\mathcal{S}_s$ contains a pure element and equals some $\mathcal{L}_{s,\sigma}$.

    As for $(i)$, since $v(x^2)=v(x)^2$, $\mu_2$ maps $[a\,|\,b]$ to $[a^2\,|\,b^2]$; thus, $\mu_2$ is linear and $\mu_2(pw)=p^2\mu_2(w)$. As $e_s^2=e_s$, this gives $\mu_2(\mathcal{S}_s)=e_s\mu_2(\mathcal{S})$, so $\mathcal{S}$ is invariant under $\mu_2$ iff every $\mathcal{S}_s$ is. This always holds for $\{\60\}$ and $\mathcal{I}_s\times\mathcal{I}_s$. As $\mu_2(cw)=c^2\mu_2(w)$, a line $\mathcal{I}_sw$ is invariant iff $\mu_2(w)\in\mathcal{I}_sw$, i.e., iff $u^2=cu$ and $v^2=cv$ for some $c\in\mathcal{I}_s$. This holds with $c=u$ if $v=\60$ or $u=v$, and with $c=v$ if $u=\60$. Otherwise, $u,v\neq\60$ and $u\neq v$, whereas cancelling in the field $\mathcal{I}_s$ gives $c=u=v$. Hence, a line is invariant iff $w$ is pure, and $(i)$ is equivalent to~\eqref{eq:frob_lines}.

    Finally, $(ii)$ and $(iii)$ are equivalent by~\cite{xyz_codes}, since pure generators of the three types form the blocks of~\eqref{eq:PCM_XYZ}, and~\eqref{eq:prod_XYZ} expresses their commutation. Choosing $\rs(\6H_\sigma)=\mathcal{W}_\sigma$, which is cyclic because the cyclic shift preserves Pauli types, the component codes $\code{C}_\sigma = \mathcal{W}_\sigma^\perp$ are cyclic as well (Section~\ref{subsec:cyclic_codes}).
\end{IEEEproof}

\begin{Cor}[Completeness of the cyclotomic construction]
\label{cor:frobenius_completeness}
   Let $n$ be odd and let $\mathcal{S}$ satisfy the equivalent conditions of Proposition~\ref{prop:frobenius}, with $\mathcal{W}_\sigma\neq\{\60\}$ for every $\sigma$.  Then $\mathcal{S}$ arises from Proposition~\ref{prop:cyc_construction} iff the $\mathcal{W}_\sigma$'s have trivial pairwise intersections.
\end{Cor}

\begin{IEEEproof}
    With $\mathcal{S}=\bigoplus_s\mathcal{S}_s$ as in the proof of Proposition~\ref{prop:frobenius}, the $\mathcal{W}_\sigma$'s intersect pairwise trivially iff $\dim_{\mathcal{I}_s}\mathcal{S}_s\leq1$ for every $s$. 
    Indeed, if $\mathcal{S}_s=\mathcal{I}_s\times\mathcal{I}_s$, then $[c\,|\,\60]$, $[\60\,|\,c]$, $[c\,|\,c]\in\mathcal{S}_s$ for every $c\in\mathcal{I}_s$, so $\mathcal{I}_s$ lies in all three $\mathcal{W}_\sigma$. 
    Conversely, if $0\neq v\in\mathcal{W}_\sigma\cap\mathcal{W}_\tau$ with $\sigma\neq\tau$, then $e_sv\neq0$ for some $s$, and the corresponding pure $\sigma$- and $\tau$-type elements of $\mathcal{S}_s$ are linearly independent over $\mathcal{I}_s$, so $\mathcal{S}_s=\mathcal{I}_s\times\mathcal{I}_s$.
    Now let $\dim_{\mathcal{I}_s}\mathcal{S}_s\leq1$ for every $s$. Each nonzero $\mathcal{S}_s$ then consists of pure elements of a single Pauli type; let $A_\sigma$ index those of $\sigma$-type, $f_\sigma=\prod_{s\in A_\sigma}M_s$ and $h_\sigma=(x^n-1)/f_\sigma$. Each such $\mathcal{S}_s$ contains all pure $\sigma$-type elements supported on $\mathcal{I}_s$, so $\mathcal{W}_\sigma=\bigoplus_{s\in A_\sigma}\mathcal{I}_s=\langle h_\sigma\rangle$, and $f_\sigma\neq1$ as $\mathcal{W}_\sigma\neq\{\60\}$. The sets $A_\sigma$ are pairwise disjoint, giving~\eqref{eq:cond_coprime}, and the pairwise orthogonality of the check spaces, $(x^n-1)\mid h_\sigma\tilde h_\tau$, is equivalent to~\eqref{eq:cond_orth} since $x^n-1$ is squarefree. Hence, $f_{\mathrm X},f_{\mathrm Y},f_{\mathrm Z}$ satisfy Definition~\ref{def:cyclic_XYZ}, and $\mathcal{S}$ arises from Proposition~\ref{prop:cyc_construction} by Corollary~\ref{cor:cyc_converse}. 
    Conversely, the construction yields cyclic stabilizers generated by pure elements, hence invariant under $\mu_2$ by Proposition~\ref{prop:frobenius}, whose check spaces intersect pairwise trivially by Proposition~\ref{prop:cyc_params}.
\end{IEEEproof}

\section{Explicit Instances}
\label{sec:examples}

We now give explicit instances of cyclic XYZ codes.  We first characterize the lengths for which such codes exist.

\begin{Cor}[Existence of cyclic XYZ codes]
\label{cor:admissible}
    A cyclic XYZ code with odd length $n$ and $k\geq1$ exists iff there are at least three nonzero negation orbits or exactly two in which at least one is paired.
\end{Cor}

\begin{IEEEproof}
    By Proposition~\ref{prop:cyc_construction} and Corollary~\ref{cor:cyc_converse}, cyclic XYZ codes correspond to three nonempty, pairwise disjoint sets of negation orbits $\Omega_\sigma$ with the choices~\eqref{eq:coset_choice}. By~\eqref{eq:k_cosets}, $k$ is the number of elements of $\mathbb{Z}_n$ not covered by $\mathcal{T}_{\mathrm X}\cup\mathcal{T}_{\mathrm Y}\cup\mathcal{T}_{\mathrm Z}$. 
    With at least three nonzero orbits, assigning one to each type leaves $0$ uncovered, i.e., $M_0\mid g$, so $k\geq1$. 
    With two nonzero orbits $\mathcal{O}_a$ and $\mathcal{O}_b$, where $\mathcal{O}_a$ is paired, assigning $\mathcal{O}_0$, $\mathcal{O}_a$, $\mathcal{O}_b$ to different types with $\mathcal{T}_{\sigma,a}=\mathcal{C}_a$ leaves $\mathcal{C}_{-a}$ uncovered.
    
    Conversely, three nonempty disjoint sets of orbits require at least three orbits, hence at least two nonzero ones, as $\mathcal{O}_0 = \{0\}$ is a single orbit. 
    If there are exactly two, each of the three orbits forms one $\Omega_\sigma$. 
    If, moreover, both nonzero orbits are self-negating, \eqref{eq:coset_choice} forces every orbit, including $\mathcal{O}_0$, to be covered entirely, so $k=0$.
\end{IEEEproof}

The exact minimum distances of the codes described below were computed using the decision-tree method of~\cite{Ott2025_decision_tree}.

\begin{Exa}[The \emph{reversible} code $\code{C}_{\text{cyc}} \llbracket 43, 1, 13 \rrbracket$]
\label{exa:code43}
    For $n = 43$, we get the following cyclotomic cosets, closed under negation,
    \begin{equation}
    \label{eq:cosets43}
    \begin{array}{@{}r@{\;=\;}l@{}}
        \mathcal{C}_1 & \{\pm 1, \pm 2, \pm 4, \pm 8, \pm 11, \pm 16, \pm 21\}, \\
        \mathcal{C}_3 & \{\pm 3, \pm 5, \pm 6, \pm 10, \pm 12, \pm 19, \pm 20\}, \\
        \mathcal{C}_7 & \{\pm 7, \pm 9, \pm 13, \pm 14, \pm 15, \pm 17, \pm 18\},
    \end{array}
    \end{equation}
    so that $x^{43 } - 1 = M_0 M_1 M_3 M_7$ with $M_1, M_3, M_7$ self-reciprocal of degree $14$, and the four orbits are $\mathcal{O}_0 = \{ 0 \}$ and $\mathcal{O}_s = \mathcal{C}_s$, with $s \in \{ 1, 3, 7 \}$.
    Choosing $\alpha$ as a root of $M_1 = x^{14} + x^{13} + x^{11} + x^{7} + x^{3} + x + 1$, we also have
    \begin{equation}
    \label{eq:minpoly43}
    \begin{array}{@{}r@{\;=\;}l@{}}
        M_3 & x^{14}{+}x^{12}{+}x^{10}{+}x^{7}{+}x^{4}{+}x^{2}{+}1, \\
        M_7 & x^{14}{+}x^{11}{+}x^{10}{+}x^{9}{+}x^{8}{+}x^{7}{+}x^{6}{+}x^{5}{+}x^{4}{+}x^{3}{+}1\rlap{.}
    \end{array}
    \end{equation}
    Setting $\mathcal{T}_{\mathrm X}=\mathcal{C}_1$, $\mathcal{T}_{\mathrm Y}=\mathcal{C}_7$, and $\mathcal{T}_{\mathrm Z}=\mathcal{C}_3$ yields a reversible Pauli-symmetric code, since multiplication by $t=6$ cyclically permutes the three defining sets. 
    Since $k=\deg g=1$ forces $g=M_0$, the three nonzero orbits must be assigned to distinct Pauli types. 
    Hence, up to a relabeling of the Pauli types, this is the only cyclic XYZ code of length $43$ with $k=1$.
    The classical component codes are $\code{C} [43, 29, 6]$, and each underlying \ac{CSS} code is $\code{C} \llbracket 43, 15, 6 \rrbracket$.
\end{Exa}

\begin{Exa}[The \emph{non-reversible} code $\code{C}_{\text{cyc}} \llbracket 31, 16, 5 \rrbracket$]
\label{exa:code31split}
    For $n=31$, the cyclotomic cosets modulo $31$ are $\mathcal{C}_0 = \{0\}$ and
    \begin{equation}
    \label{eq:cosets31nr}
    \begin{array}{@{}r@{\;=\;}l@{\quad \,\,\,}l@{\;=\;}l@{}}
        \mathcal{C}_1 & \{1, 2, 4, 8, 16\},   & \mathcal{C}_{15} & \{15, 23, 27, 29, 30\}, \\
        \mathcal{C}_3 & \{3, 6, 12, 17, 24\}, & \mathcal{C}_7    & \{7, 14, 19, 25, 28\}, \\
        \mathcal{C}_5 & \{5, 9, 10, 18, 20\}, & \mathcal{C}_{11} & \{11, 13, 21, 22, 26\}.
    \end{array}
    \end{equation} 
    There are therefore four orbits: $\mathcal{O}_0 =\{0\}$ and the three paired orbits $\mathcal{O}_1 = \mathcal{C}_1 \cup \mathcal{C}_{15}$, $\mathcal{O}_3 = \mathcal{C}_3 \cup \mathcal{C}_7$, $\mathcal{O}_5 = \mathcal{C}_5 \cup \mathcal{C}_{11}$.
    Choosing $\alpha\in\mathbb{F}_{32}$ as a root of $M_1 = x^5 + x^2 + 1$, the minimal polynomials are $M_0 = x + 1$ and
    \begin{equation}
    \label{eq:minpoly31}
    \begin{array}{@{}r@{\;=\;}l@{\quad}r@{\;=\;}l@{}}
        M_1 & x^5{+}x^2{+}1,             & M_7    & x^5{+}x^3{+}x^2{+}x{+}1, \\
        M_3 & x^5{+}x^4{+}x^3{+}x^2{+}1, & M_{11} & x^5{+}x^4{+}x^3{+}x{+}1, \\
        M_5 & x^5{+}x^4{+}x^2{+}x{+}1,   & M_{15} & x^5{+}x^3{+}1.
    \end{array}
    \end{equation}
    We assign $\mathcal{O}_1$, $\mathcal{O}_3$, $\mathcal{O}_5$ to $\sym{X}$, $\sym{Y}$, $\sym{Z}$ and split each of them, retaining $\mathcal{T}_{\mathrm{X}} = \mathcal{C}_1$, $\mathcal{T}_{\mathrm{Y}} = \mathcal{C}_7$ and $\mathcal{T}_{\mathrm{Z}} = \mathcal{C}_5$, i.e., $f_{\mathrm{X}} = M_1$, $f_{\mathrm{Y}} = M_7$ and $f_{\mathrm{Z}} = M_5$, which gives a non-reversible code. 
    It is easily verified that the code is Pauli-symmetric with $t = 5$, so the Pauli-type symmetry survives even though splitting the orbits destroys reversibility. 
    Then, $g = M_0 M_3 M_{11} M_{15}$, and~\eqref{eq:k_cosets} gives $k = 31 - 3 \cdot 5 = 16$.
    The classical component codes $\code{C}_\sigma=\langle\tilde f_\sigma\rangle$ are $\code{C}_{\mathrm{X}} = \langle M_{15}\rangle$, $\code{C}_{\mathrm{Y}} = \langle M_3\rangle$ and $\code{C}_{\mathrm{Z}} = \langle M_{11}\rangle$, as $\tilde M_s=M_{-s}$, with  parameters $[31,26,3]$.
    Then, each underlying \ac{CSS} code is $\code{C} \llbracket 31, 21, 3 \rrbracket$.
\end{Exa}

\begin{remark}
\label{rem:compare}
    We want to remark that $\code{C}_{\text{cyc}} \llbracket 43, 1, 13 \rrbracket$ code attains the best known minimum distance $d=13$ of stabilizer codes with $(n,k)=(43,1)$, held by a cyclic code in~\cite{Grassl_table}, which coincides with ours up to a relabeling of the Pauli types.\footnote{Among $\mu_2$-invariant codes, this is forced. 
    For $n=43$, $\mathcal{S}=\mathcal{S}_0\oplus\mathcal{S}_1\oplus\mathcal{S}_3\oplus\mathcal{S}_7$ with $\dim\mathcal{S}_0\leq2$, and each $\mathcal{S}_s$, $s\neq0$, is $\{\60\}$ or a line of binary dimension $14$, since $\mathcal{C}_s=-\mathcal{C}_s$ makes $\mathcal{I}_s\times\mathcal{I}_s$ non-isotropic. 
    As $\dim\mathcal{S}=42=3\cdot14$, all three are lines, of pure type if $\mathcal{S}$ is $\mu_2$-invariant (Proposition~\ref{prop:frobenius}). 
    Two lines of the same type would make the code \ac{CSS} up to relabeling, with $d\leq11$ by the \ac{CSS} \ac{LP} bound below.  
    Hence, up to relabeling, Example~\ref{exa:code43} is the only $\mu_2$-invariant cyclic $\llbracket43,1\rrbracket$ code with $d>11$.}
    Likewise, $\code{C}_{\text{cyc}} \llbracket 31, 16, 5 \rrbracket$ matches the parameters of the extended \ac{QC} code $\llbracket 31, 16, 5 \rrbracket$ in~\cite{Grassl_table},    and a cyclic code with the same parameters is listed in~\cite{GrasslCyc}. 
    Moreover, both examples beat \emph{every} \ac{CSS} code with the same $(n,k)$: the \ac{LP} bound~\cite{Delsarte1973, Shor1997_MacWilliams, Rains1998_enumerators, Ashikhmin1999_bounds}, applied in exact arithmetic to the weight enumerators of the $\sym{X}$- and $\sym{Z}$-type check spaces of arbitrary \ac{CSS} codes,  imposing orthogonality and distance constraints over all splits of the $n-k$ independent stabilizer generators, yields $d\leq11$ for \ac{CSS} $\llbracket43,1\rrbracket$ codes and $d\leq4$ for \ac{CSS} $\llbracket31,16\rrbracket$ codes~\cite{Sarvepalli2009_asymmetric}.

    We summarize the above results in Table~\ref{tab:code_parameters}.
\end{remark}.

We verify by using the tool available in~\cite{wang2026} that the introduced cyclic code instances are genuine for Definition~\ref{def:genuine_XYZ}.

\begin{table}[t]
\centering
\caption{Parameters of proposed Cyclic XYZ Codes, compared to the underlying \ac{CSS} Codes, the best known stabilizer codes and the \ac{LP} upper-bound for the minimum distance of \ac{CSS} codes.}
\resizebox{\columnwidth}{!}{
\begin{tabular}{c c c c}
    Cyclic & Underlying & Best stabilizer & Best CSS \\
    XYZ Code & \ac{CSS} Codes & (known so far) & (LP bound) \\
    \hline
    $\llbracket 43, 1, 13 \rrbracket$ & $\llbracket 43, 15, 6 \rrbracket$ & $\llbracket 43, 1, 13 \rrbracket$ & $d \leq 11$ \\
    $\llbracket 31, 16, 5 \rrbracket$ & $\llbracket 31, 21, 3 \rrbracket$ & $\llbracket 31, 16, 5 \rrbracket$ & $d \leq 4$  
\end{tabular}
}
\label{tab:code_parameters}
\end{table}

\section{Conclusions}
\label{sec:conclusions}

We have introduced cyclic XYZ codes, whose classical component codes are cyclic and whose three Pauli types use disjoint sets of irreducible factors of $x^n-1$. This fixes the dimension and reduces orthogonality between each pair of \acp{PCM} to a coprimality condition; cyclotomic cosets describe the whole family.
By fixing the length and the dimension, the two presented codes attain the best known minimum distance of stabilizer codes, exceed those of the underlying \ac{CSS} codes, and beat the \ac{LP} upper bound for \ac{CSS} codes.

\clearpage

\IEEEtriggeratref{14}
\bibliographystyle{IEEEtran}
\bibliography{strings, Archive_short}

\end{document}